\documentclass[%
 reprint,
superscriptaddress,
footinbib,
 amsmath,amssymb,
 aps,
 showkeys,
]{revtex4-1}
\usepackage{graphicx}% Include figure files
\usepackage{dcolumn}% Align table columns on decimal point
\usepackage{bm}% bold math
\usepackage{hyperref}
\usepackage{footnote}
\usepackage{xcolor}
\usepackage[mathlines]{lineno}
\usepackage{fontenc}
\usepackage{graphicx}
\usepackage{dcolumn}
\usepackage{bm}
\usepackage{booktabs}
\usepackage{array,multirow}
\usepackage{amsmath}
\usepackage{amsfonts}
\usepackage{amssymb}
\usepackage{mathrsfs}
\usepackage{enumerate}
\usepackage{fancyhdr}
\usepackage{xcolor}
\usepackage{graphicx}
\usepackage{listings}
\usepackage{float}
\usepackage{verbatim}
\usepackage{braket}
\usepackage{centernot}
\usepackage{setspace}
\usepackage[normalem]{ulem}
\usepackage{wrapfig}
\usepackage{multirow}
\usepackage{comment}
\usepackage{kantlipsum}
\usepackage[bottom]{footmisc}
\allowdisplaybreaks
\usepackage{lipsum, babel}
\hypersetup{breaklinks = true, colorlinks = true, citecolor = blue, linkcolor = blue, urlcolor = blue}
\usepackage[mathlines]{lineno}
\begin{document}
\preprint{APS/123-QED}

\title{Entanglement as a probe of hadronization}
\author{Jaydeep Datta}
\email{jaydeep.datta@stonybrook.edu}
\affiliation{Center for Nuclear Frontiers in Nuclear Science, Department of Physics and Astronomy, Stony Brook University, New York 11794-3800, USA}
\author{Abhay~Deshpande}
\email{abhay.deshpande@stonybrook.edu}
\affiliation{Center for Nuclear Frontiers in Nuclear Science, Department of Physics and Astronomy, Stony Brook University, New York 11794-3800, USA}
\affiliation{Department of Physics, Brookhaven National Laboratory, Upton, New York 11973-5000, USA}
\author{Dmitri E.~Kharzeev}
\email{dmitri.kharzeev@stonybrook.edu}
\affiliation{Center for Nuclear Theory, Department of Physics and Astronomy, Stony Brook University, New York 11794-3800, USA}
\affiliation{Energy and Photon Sciences Directorate, Condensed Matter and Materials Sciences Division, Brookhaven National Laboratory, Upton, New York 11973-5000, USA}
\author{Charles Joseph Naïm}
\email{charlesjoseph.naim@stonybrook.edu}
\affiliation{Center for Nuclear Frontiers in Nuclear Science, Department of Physics and Astronomy, Stony Brook University, New York 11794-3800, USA}
\author{Zhoudunming~Tu}
\email{zhoudunming@bnl.gov}
\affiliation{Department of Physics, Brookhaven National Laboratory, Upton, New York 11973-5000, USA}

\date{\today}

\begin{abstract} 
Recently, it was discovered that the proton structure at high energies exhibits maximal entanglement. This leads to a simple relation between the proton’s parton distributions and the entropy of hadrons produced in high-energy inelastic interactions, that has been experimentally confirmed.   
In this Letter, we extend this approach to the production of jets. Here, the maximal entanglement predicts a relation between the jet fragmentation function and the entropy of hadrons produced in jet fragmentation. We test this relation using the ATLAS Collaboration data on jet production at the Large Hadron Collider, and find a good agreement between the prediction based on maximal entanglement within the jet and the data. 
This study represents the first use of quantum entanglement framework in experimental study of the hadronization process, offering a new perspective on the transition from perturbative to non-perturbative QCD. Our results open the door to a more comprehensive understanding of the quantum nature of hadronization.
\end{abstract}
\maketitle
%\tableofcontents
\textit{Introduction.} In Quantum Chromodynamics (QCD), the asymptotic freedom implies that quarks and gluons (partons) interact weakly at short distances. At large distances, parton interactions become strong and result in color confinement that  
binds partons within hadrons. Particles inside a high-energy hadron are typically described by parton distribution functions (PDFs), which represent the probability of finding a parton with a given momentum (Bjorken $x$) and resolution scale ($Q^2$). 
% fraction of the hadron's energy (Bjorken $x$) and a specific momentum transfer $Q^2$.
These distribution functions are  probabilistic in nature, and do not contain full information about the proton wave function that represents a coherent quantum superposition of Fock states with different number of partons, and with different phases. 

Because of the uncertainty relation between the phase and occupation number in quantum mechanics, they cannot be both measured precisely. In high energy interactions, the phases of the Fock components cannot be measured due to the short time of the interaction \cite{Kharzeev:2017qzs, kharzeev2022quantum}. The measured density matrix of the proton thus represents a mixed density matrix obtained from the initial pure density matrix by tracing over the unobserved phases. This mixed density matrix is an incoherent superposition of states with fixed numbers of partons, and the resulting probabilistic description corresponds to the parton model \cite{kharzeev2022quantum}.

As a result of QCD evolution, the {\it maximal entanglement} inside a hadron is reached at small $x$ as a result of (almost) equal probabilities $p_n$ of a large number of configurations with different number of partons $n$ \cite{Kharzeev:2017qzs}. In fact, this maximal entanglement occurs only within a subspace of the full Hilbert space of the proton states which is described by a relativistic field theory -- QCD -- and is thus infinite. If the maximal entanglement occurs within a Hilbert subspace of dimension $N$, then $p_n =1/N$, and the von Neumann entanglement entropy is $S_E = - \sum_n p_n \ln p_n = \ln N$ which for large $N$ results in a simple Boltzmann-like relation $S_E = \ln {\bar N}$, where ${\bar N}= \sum_n p_n n$ is the value of the PDF at a given momentum fraction, $x$, and momentum transfer squared, $Q^2$. By generalizing the concept of local parton-hadron duality \cite{azimov1985similarity} to the number of produced particles per event, also known as multiplicity distributions, we find a relationship between the parton distribution functions and the entropy of the produced hadrons.

Strong experimental evidence for the maximal entanglement at small $x$ has been observed in various QCD processes, including minimum-bias proton-proton ($pp$) collisions at the LHC~\cite{Tu:2019ouv}, inclusive electron-proton ($ep$) deep inelastic scattering (DIS) \cite{Kharzeev:2021yyf,Hentschinski:2021aux,Hentschinski:2022rsa,Hentschinski:2024gaa}, and diffractive DIS \cite{Hentschinski:2023izh}. Recently, the QCD evolution of entanglement entropy has been determined and validated against hadron entropy measured in different rapidity windows~\cite{Hentschinski:2024gaa}. These studies have established a quantum entanglement approach to describing the QCD structure of hadrons, providing a complementary method for studying parton distributions inside hadrons. For other related work, see Refs.~\cite{Kutak:2011rb,Peschanski:2012cw,Armesto:2019mna,Neill:2018uqw,Armesto:2019mna,Kovner:2018rbf,Chachamis:2023omp,Liu:2022hto,Liu:2023eve,Liu:2022bru,Liu:2022qqf,Stoffers:2012mn,Asadi:2023bat,Kou:2022dkw,Kutak:2023cwg,Dumitru:2023qee,Ehlers:2022oke,Ehlers:2022oal,Florio:2024aix,Grieninger:2023knz,Grieninger:2023pyb,Ikeda:2023zil,gursoy2024universal,Berges:2017hne,Berges:2017zws,Dumitru:2023fih,Dumitru:2022tud,Ramos:2022gia,Moriggi:2024tbr,Ramos:2020kaj,Bhattacharya:2024sno,Hatta:2024lbw}.

In this Letter, we extend the test of maximal entanglement to hadronization of high-momentum jets. At high momentum, the QCD renormalization group flow turns a jet into a complicated multi-parton system, that can be expected to exhibit the maximal entanglement. Indeed, the parton distribution functions and fragmentation functions (FF), which describe how a parton fragments into hadrons, are related by crossing symmetry, which gives rise to reciprocal relationships between these quantities \cite{Gribov:1971zn}. It is thus natural to expect that the jet state is also maximally entangled, giving rise to the relation between the jet fragmentation function and the entropy of produced hadrons. Recent quantum simulations of hadronization in jet production~\cite{Florio:2023dke,Florio:2024aix} probed the real-time dynamics of the entanglement entropy production process, but the relation between entanglement entropy and fragmentation function has not yet been explored.  

\begin{figure}
    \centering
    \includegraphics[scale=0.38]{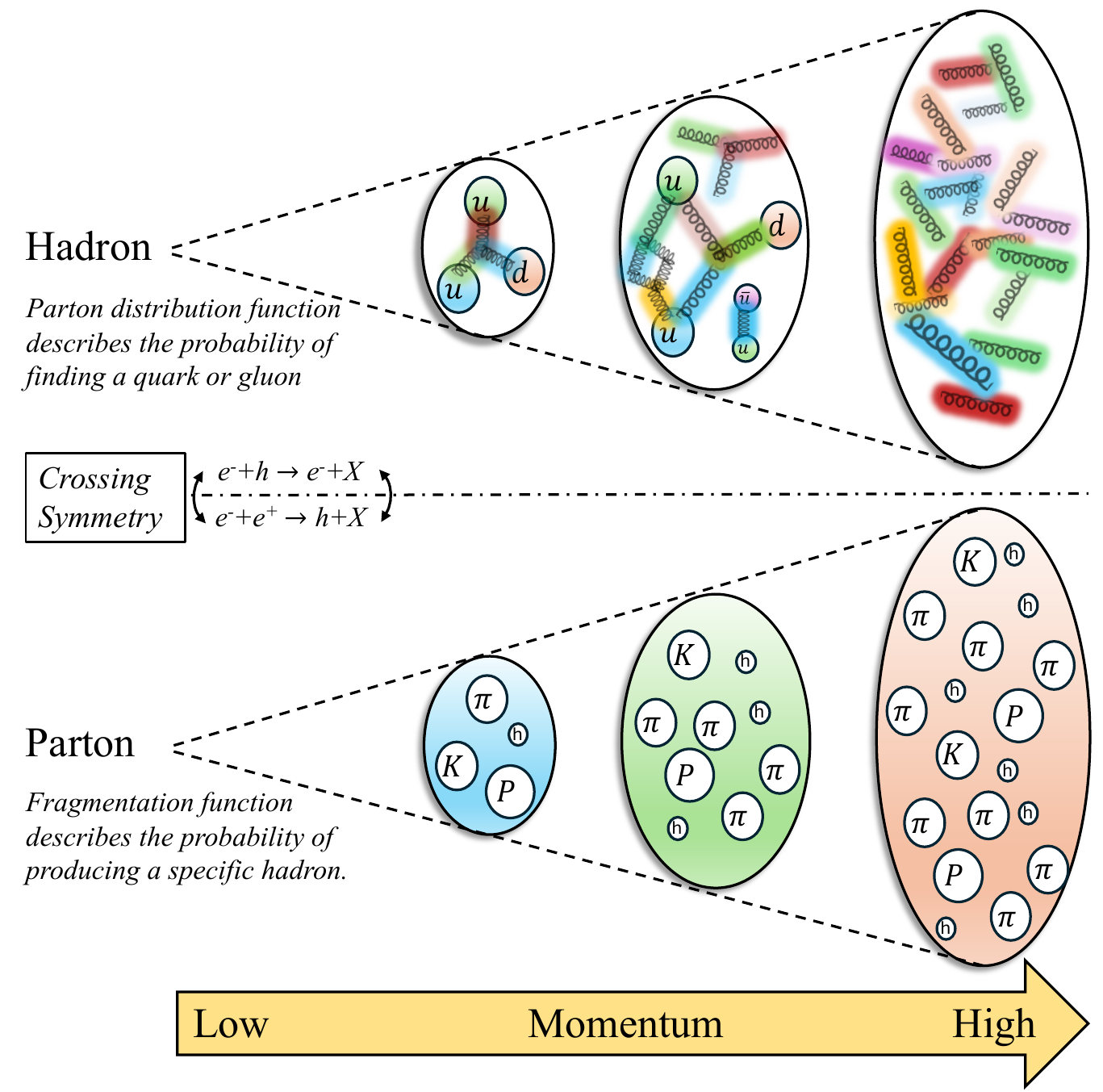}
\caption{Illustration of hadron and parton from low to high momentum and their relations in terms of parton distribution function and fragmentation function.}
    \label{fig:figure_0}
\end{figure}

Specifically, we extend the universal entanglement entropy model~\cite{Hentschinski:2024gaa}, an improved version of the original KL model~\cite{Kharzeev:2017qzs}, to relate the jet fragmentation function to the entropy of the final hadron state. Within this approach, we compare the entanglement entropy derived from FFs with the hadron entropy calculated from the charged hadron multiplicity distribution within jets, as measured by the ATLAS experiment~\cite{ATLAS:2011eid,ATLAS:2016vxz,ATLAS:2019rqw} at the LHC. Additionally, we investigate hadron entropy inside jets at the upcoming EIC and compare Monte Carlo simulations using PYTHIA~8~\cite{pythia8} to the entanglement entropy of FFs. Some details of this work are presented in the Supplemental Material~\footnote{See Supplemental Material at \url{url_link_added_by_journal} for details on how to match kinematic variables between data and models, uncertainties on the entanglement entropy model, and uncertainties associated with fragmentation functions, which includes Refs.~\cite{ATLAS:2011myc,Moffat:2021dji,Bertone:2018ecm} }.

\textit{Theory.}
In quantum field theory, scattering amplitudes are analytic in the Mandelstam plane, defined by the kinematic variables $s$, $t$, and $u$. Here the variables $s$, $t$, and $u$ correspond to the reaction in ``$s$, $t$, $u$ channel", respectively, for $2\rightarrow 2$ processes, which results in the crossing symmetry relating their amplitudes~\cite{PhysRev.96.1433}. For example, the high energy DIS process $e^{-}+h\rightarrow e^{-}+X$ is related to  electron-positron annihilation $e^{-}+e^{+}\rightarrow h+X$. As a result, the roles that PDFs and FFs play are very similar; see Fig.~\ref{fig:figure_0} for an illustration. 

In the parton fragmentation process, however, the initial von Neumann entropy is not zero and is determined by the entanglement created in the production of the parton  pair, which is determined by the dimensionality of Hilbert space describing the color states of the pair, see e.g.~\cite{Florio:2024aix}. 
Specifically, assuming a ``bare" parton-antiparton initial configuration, we get $S_{q}=\ln{(N_c)}$ and $S_{g}=\ln{(N^{2}_{c}-1)}$, where $N_c=3$ represents the number of colors in the $SU(3)$ gauge group. A similar calculation can be found here~\cite{Miller:2003ci}.

Therefore, we present an approach that connects the entropy of charged hadrons within jets to their fragmentation functions for a maximally entangled hadronization process as follows:
\begin{equation}
    S^{q/g}_{\text{FF}} = S_{q/g} + \ln \left[ \int_{z_{\text{min}}(p_{\perp}^{\mathrm{jet}})}^{1} d z \, D_{q/g}^{\text{h}}\left(z, \mu^2 \right) \right].
\label{eq:model}
\end{equation}
\noindent 
Here $D^{h}_{q/g}$ is the FF of charged hadron from either a quark or gluon. The second term describes the entropy of the fragmentation process using the FF with $\mu^2$ as the hard scale. 

\textit{Experimental data.} We investigate jet production in proton-proton collisions with a transverse momentum $p_{\perp}^{\text{jet}}$. A jet is initiated by a quark or gluon, which evolves over time by producing a parton shower that subsequently fragments into hadrons. 
The entropy $S$ of all hadrons in jet is related to the number of charged hadrons produced in the final state and is defined as~\cite{Kharzeev:2017qzs,Tu:2019ouv,Hentschinski:2024gaa}:
\begin{equation} S_{\text{hadrons}} = -\sum{P_{n}\ln{\left(P_{n}\right)}}, \label{eq:entropy} \end{equation}
\noindent where $P_{n}$ represents the probability of detecting $n$ charged hadrons. 

The data analyzed in this Letter are measurements of jet productions and their associated hadrons in $pp$ collisions, which were published by the ATLAS experiment at center-of-mass energies of $\sqrt{s} = 7 \, \text{TeV}$ \cite{ATLAS:2011eid} and $\sqrt{s} = 13 \, \text{TeV}$ \cite{ATLAS:2019rqw}. The kinematic phase spaces are summarized in Table.~\ref{tab:kinematics}.
\begin{figure}
    \centering
    \includegraphics[scale=0.50]{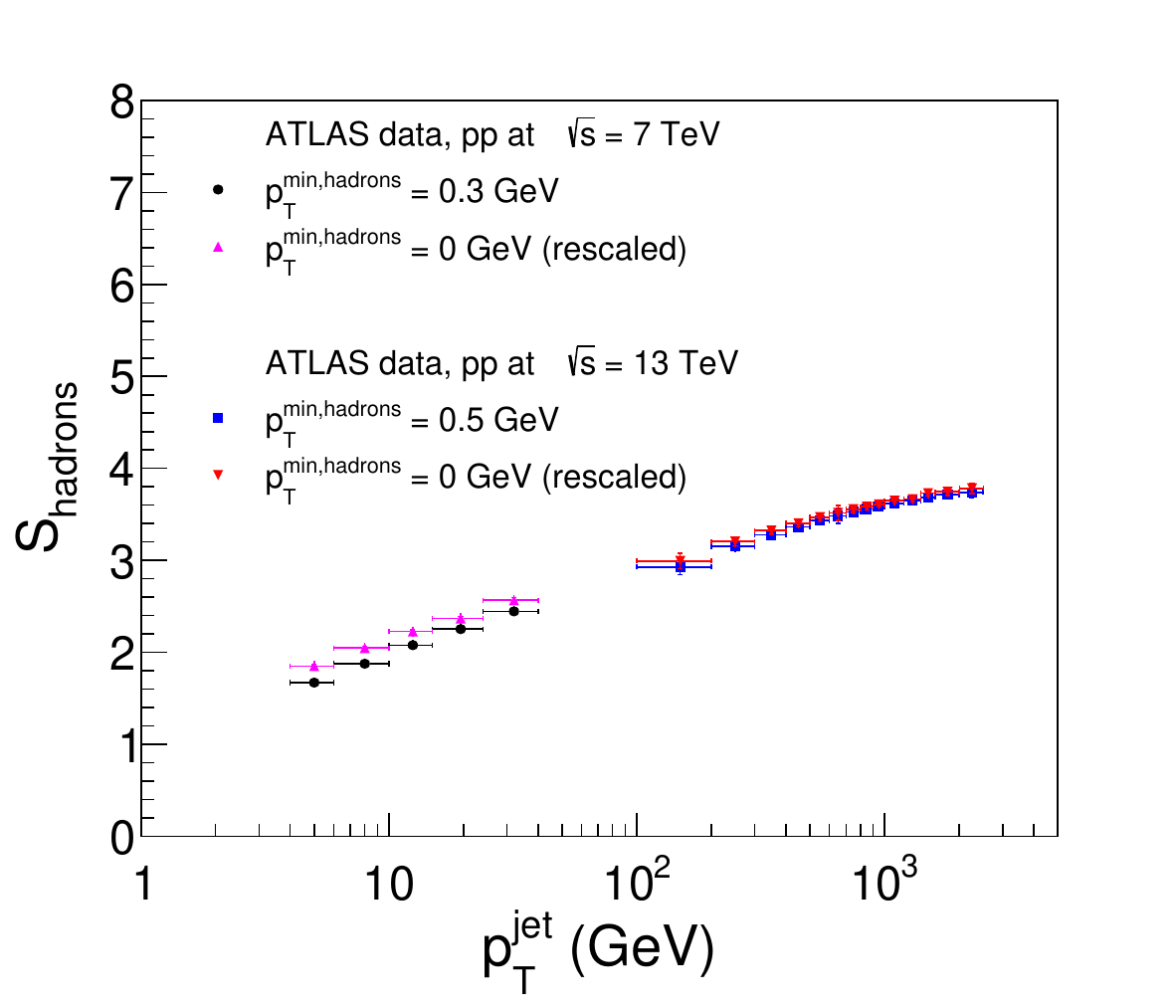}
\caption{Charged hadron entropy $S_{\text{hadrons}}$ as a function of jet transverse momentum $p_{\perp}^{\text{jet}}$ for ATLAS data at $\sqrt{s} = 7 \, \text{TeV}$ \cite{ATLAS:2011eid} and $\sqrt{s} = 13 \, \text{TeV}$ \cite{ATLAS:2019rqw}, calculated using Eq.~\ref{eq:entropy}.}
    \label{fig:S_pTjet_data}
\end{figure}

These measurements provide the charged particle multiplicity distribution in jets, $P(n)$, as a function of the jet's transverse momentum, $p_{\perp}^{\mathrm{jet}}$. Using these data, the entropy $S_{\text{hadrons}}$ in Eq.~\eqref{eq:entropy} is calculated as a function of $p_{\perp}^{\mathrm{jet}}$, as shown in Fig.~\ref{fig:S_pTjet_data}. The data reveals an increase in entropy with rising jet $p_{\perp}^{\text{jet}}$. To facilitate comparison with Eq.~\eqref{eq:model}, the data is re-expressed in terms of the mean value $\left\langle z \right\rangle$ for each $p_{\perp}^{\mathrm{jet}}$ bin. The relationship between $\left\langle z \right\rangle$ and $p_{\perp}^{\mathrm{jet}}$ is determined through a PYTHIA simulation of $pp$ collisions at the relevant center-of-mass energy. Note that the threshold on $p_{\perp}^{\text{hadrons}}$ directly influences the charged hadron multiplicity within the jet, where the entropy is extrapolated down to zero $p_{\perp}^{\text{hadrons}}$ for both dataset. Before and after the rescaling are both shown in Fig.~\ref{fig:S_pTjet_data}. Further details on the mapping between $\left\langle z \right\rangle$ and $p_{\perp}^{\mathrm{jet}}$ and how to scale the data down to $p_{\perp}^{\text{hadrons}}=0$ are provided in the Supplemental Material.
\begin{table}[h!]
    \centering
    \begin{tabular}{|c|c|c|c|}
        \hline
        $\sqrt{s}$ (TeV) &  $|y|$ range & $p_{\perp}^{\mathrm{jet}}$ range (GeV) & $p_{\perp}^{\text{hadrons}}$ (GeV) \\ \hline
        $7$  & $< 1.9$  & $4 < p_{\perp}^{\mathrm{jet}} < 40$  &  $>$ 0.3  \\ \hline
        $13$ & $< 2.1$  & $100 < p_{\perp}^{\mathrm{jet}} < 2500$ & $>$ 0.5  \\ \hline
    \end{tabular}
    \caption{\label{tab:kinematics}Summary of the kinematic phase space for charged hadron multiplicity distribution in jets in $pp$ collisions at the ATLAS experiment \cite{ATLAS:2011eid,ATLAS:2019rqw}.}
    \label{tab:ATLAS_kinematic_phase_space}
\end{table}

\begin{figure*}[tbh]
    \centering
   \includegraphics[scale=.9]{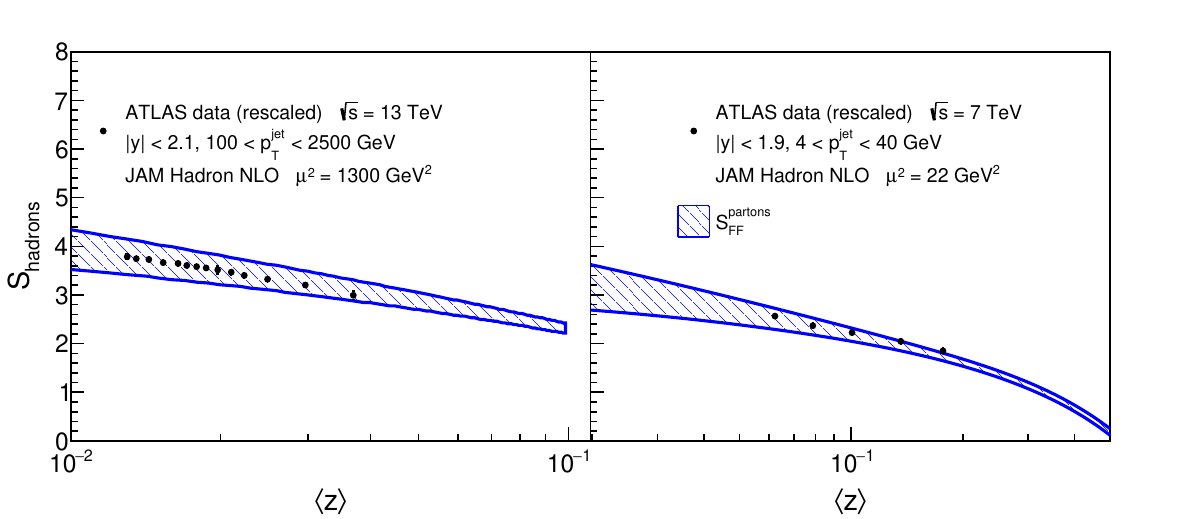}
\caption{The entropy $S_{\text{hadrons}}$ as a function of $\left\langle z \right\rangle$ for $S_{\text{FF}}^{\text{partons}}$ — incorporating gluons, $u$-(anti)quarks, and $d$-(anti)quarks — is shown using JAM fragmentation functions at NLO for $\mu^2 = 1300$ GeV$^2$, compared with ATLAS data at $\sqrt{s} = 13$ TeV \cite{ATLAS:2019rqw} (left). Additionally, the results at $\mu^2 = 22$ GeV$^2$ are compared with ATLAS data at $\sqrt{s} = 7$ TeV \cite{ATLAS:2011eid} (right). The uncertainties are calculated at the 1$\sigma$ level. The total entropy $S_{\text{FF}}^{\text{partons}}$ is derived from the sum of the individual entropies of each parton, with each contribution normalized by the average fraction of jets produced by that parton from a PYTHIA simulation.}
\label{fig:JAM_ATLAS_data}
\end{figure*}

\textit{Results.} The calculations presented in this section are obtained using Eq.~\ref{eq:model} with the JAM \cite{Moffat:2021dji} and NNFF1.1 \cite{Bertone:2018ecm} fragmentation functions using of all hadrons at the Next-to-Leading Order (NLO) for quarks and gluons. The hard scale $\mu$ of the fragmentation function is set to the typical transverse momentum of charged hadrons within the jet, as the relevant scale relates to jet fragmentation into hadrons rather than jet production. Uncertainties are computed using all FF sets, with the error band representing a 1 standard deviation ($\sigma$) confidence level. The calculations show minor variations with $\mu$ and are predominantly influenced by these uncertainties. See Supplemental Material for details of the choice of scale and its uncertainties associated with the FFs. 

Figure~\ref{fig:JAM_ATLAS_data} left compares Eq.~\eqref{eq:model} with the JAM FF at NLO at $\mu^{2} = 1300 \, \text{ GeV}^2$ and the ATLAS data at $\sqrt{s} = 13 \, \text{TeV}$. The predicted entanglement entropy, $S^{\mathrm{partons}}_{\mathrm{FF}}$, is the weighted average entropy of light quarks (and their anti quarks) and gluons. Their relative contributions, namely the fraction of different quarks and gluon jets, are provided by the PYTHIA simulation. See the Supplemental Material for details on the evaluation of the parton jet fraction. The entropy is observed to increase in both the data and predictions as the value of $\left\langle z \right\rangle$ decreases. An excellent match is found between the data and $S_{\text{FF}}^{\text{partons}}$.

The data at $\sqrt{s} = 13 \, \text{TeV}$ and 7 TeV allows access to complementary $z$ values due to the range of accessible jet $p_{\perp}^{\text{jet}}$. Figure~\ref{fig:JAM_ATLAS_data} right compares Eq.~\eqref{eq:model} under the same conditions except for a much lower scale and the ATLAS data at $\sqrt{s} = 7 \, \text{TeV}$. These data allow probing greater values of $\left\langle z \right\rangle \sim 0.1$, thereby enabling a test of the $z$-dependence of our model. The data follows a similar trend as the calculations, and is found to be consistent with the predicted entanglement entropy. Note that towards higher $\left\langle z \right\rangle$, the relative contribution of quark and gluon jets significant changes, where gluon jets are more dominant. This agreement, again, confirms our model at lower jet transverse momentum (thus, higher $\left\langle z \right\rangle$) for 7 TeV $pp$ data. 

For data at both energies, they are consistent with the expectation of our model in this Letter. Specifically, as the jet transverse momentum increases, the quark-gluon scattering process ($qg\rightarrow qg$) starts to become significant comparing to gluon-gluon fusions ($gg\rightarrow gg$ and $gg\rightarrow q\bar{q}$). As a result, the fraction of quark jets becomes comparable to the gluon ones. On the other hand, as the the jet transverse momentum decreases for the 7 TeV data, the fraction of gluon jets increases. Although the underlying contribution from quarks and gluons significantly changes, the entire $z$ dependence of entanglement entropy and its data comparison have been captured very well by our model. We also estimate the contributions from heavy quarks, which are found to be negligible comparing to contributions from $u$, $d$ (anti)quarks, and gluons.    

One should note that we also compare the data with entanglement entropy based on NNFF1.1, where a similar description of the data has been observed. The quantitative difference, however, is purely raised from the difference in the FFs, where uncertainty becomes larger at lower $z$ region. A discussion regarding the details of the fragmentation functions is provided in the Supplemental Material. 

This observation of an agreement between our model and the data provides a good indication that the hadronization process exhibits maximum entanglement, which is similar to the case of protons at high beam energies. On one hand, perturbative QCD allows us to calculate the number of color states at the parton level shortly after the hard scattering, which can be used to calculate the entanglement entropy at the initial time. On the other hand, the process of hadronization afterwards cannot be calculated perturbatively. Instead, the entanglement entropy from FFs can be used in describing the total hadron production. Together, from perturbative to nonperturbative regimes, this quantum entanglement approach provides an independent tool for understanding the hadronization process. 

In addition, one can push the initial condition of jets to lower scales into less perturbative regimes, which would increase the entanglement entropy in the system to begin with, e,g., $S_{q/g} > \ln{N_{\Omega}}$, where $N_{\Omega}$ represents the number of accessible color states of the parton. For a realistic estimation, this might be possible by utilizing quantum simulation~\cite{Florio:2023dke, Gong:2021bcp} on quantum hardware. This is beyond the scope of this Letter, but could be a natural follow-up of this work. 

We predict hadron entropy in jets at the upcoming EIC, which will be essential for constraining the nonperturbative initial conditions of jets and their hadronization processes. Using PYTHIA~8~\cite{pythia8}, we simulated jet production in $ep$ collisions with 18 GeV electrons and 275 GeV protons, excluding detector effects. The phase space was restricted by a minimum $Q^{2}$ cut of 10 GeV$^{2}$, representing the virtuality of the photon.

\begin{figure}
    \centering
    \includegraphics[scale=0.5]{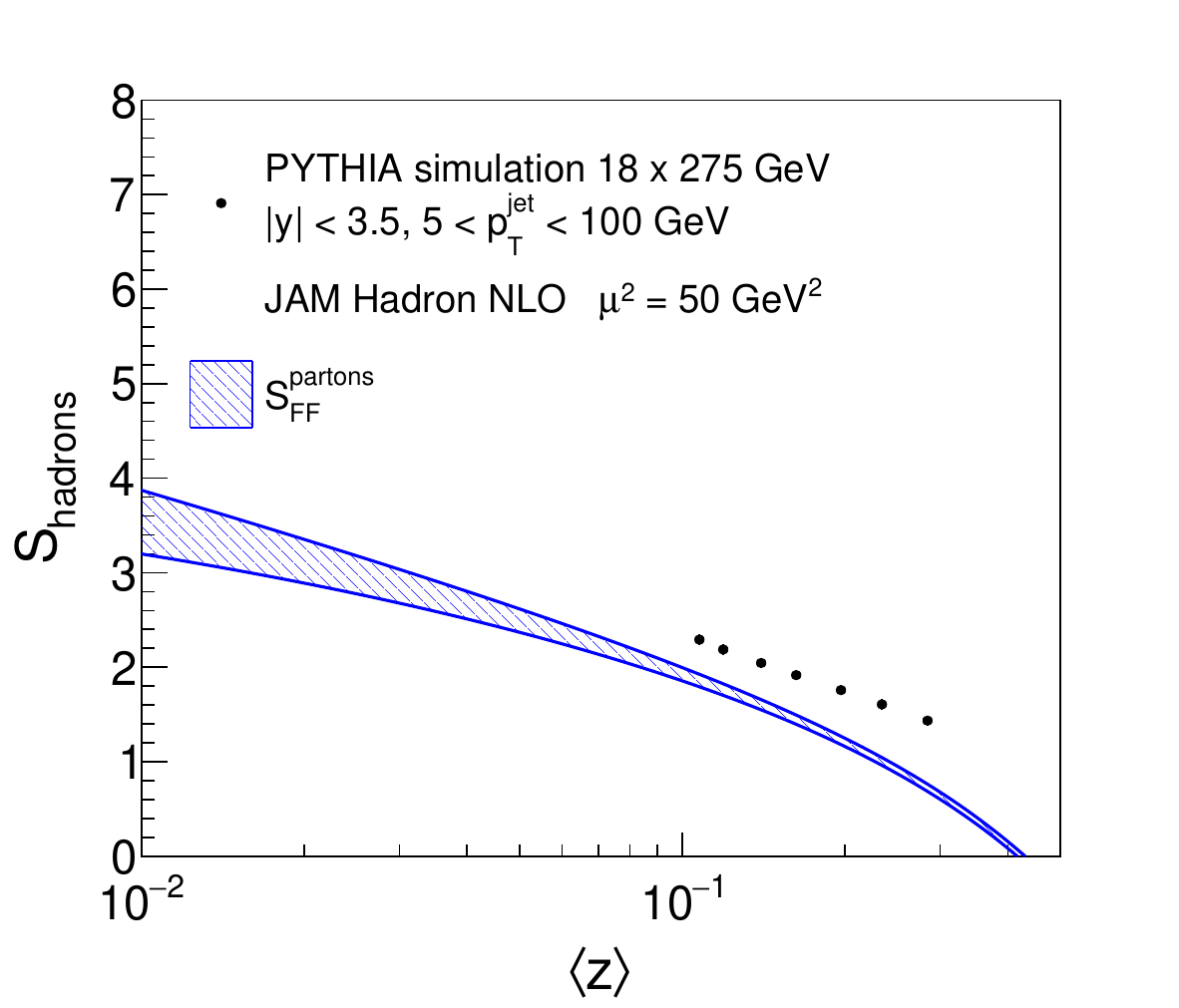}
    \caption{
    The entropy $S_{\text{hadrons}}$ as a function of $\left\langle z \right\rangle$ for $S_{\text{FF}}^{\text{partons}}$ — incorporating gluons, $u$-(anti)quarks, $d$-(anti)quarks and $c$-(anti)quarks — is shown using JAM fragmentation functions at NLO for $\mu^2 = 50$ GeV$^2$, compared to PYTHIA~8 simulation in the ePIC EIC kinematic phase space. The uncertainties are calculated at the 1$\sigma$ level. The total entropy $S_{\text{FF}}^{\text{partons}}$ is derived from the sum of the individual entropies of each parton, with each contribution normalized by the average fraction of jets produced by that parton from a PYTHIA simulation.} 
    \label{fig:JAM_NLO_hadrons_ePIC_mu2_50}
\end{figure}

Jets were selected with a radius parameter of 0.4, consistent with ATLAS criteria, and restricted to jets containing a minimum of two particles with a transverse momentum range of $5 < p_{\perp}^{\text{jet}} < 100$ GeV. To match the ePIC detector's acceptance, only jets with a pseudo-rapidity range of $-3.5 < \eta < 3.5$ were considered \cite{ABDULKHALEK2022122447}. The charged-hadron entropy in jets is plotted as a function of average value of $\left\langle z \right\rangle$ in Fig.~\ref{fig:JAM_NLO_hadrons_ePIC_mu2_50}. Note that the contribution of the $c$-(anti)quarks has been included in our calculation of the total entropy due to its significant role in jet production in $ep$ collisions. The charged hadron entropy from PYTHIA simulations is above our theoretical estimates. In $ep$ collisions, processes like $\gamma q \to gq$ and $\gamma g \to q \bar{q}$ enhance quark jet contributions at lower $\left\langle z \right\rangle$ values, reducing gluon jet contribution. This effect should lead to lower entropy values than those seen in PYTHIA simulations. EIC experiments will be essential for clarifying jet entanglement entropy, particularly in processes dominated by quark jets.

\textit{Summary.} In conclusion, we have, for the first time, explored the connection between the FFs and hadron entropy implied by the maximal entanglement. The entanglement entropy in jets is found to be related to the FF in a manner similar to its relationship with PDFs, while the initial entanglement entropy of partons has a lower bound directly tied to the number of color degrees of freedom, as expected at weak coupling dictated by the asymptotic freedom. The charged particle multiplicity distributions in jets from the ATLAS experiment at 7 TeV and 13 TeV provided an estimate of the hadron entropy associated with jet fragmentation. Our results show that the entanglement entropy, based on the FF from JAM and NNFF1.1 at the Next-to-Leading Order, agrees well with the ATLAS data, suggesting that the fragmentation process exhibits maximal entanglement for high transverse momentum jets—similar to the behavior of the proton wave function at high energies. This approach could be extended to study lower transverse momentum jet fragmentation at the upcoming Electron-Ion Collider, offering deeper insights into the nonperturbative aspects of hadronization. Exploring this direction, along with first-principle quantum simulations on quantum and classical hardware, presents significant opportunities for future research. 

\begin{acknowledgments}
The authors would like to thank Martin Hentschinski and Krzysztof Kutak for fruitful discussions on this and related topics. J.~Datta was suppoted by the Center for Frontiers in Nuclear Science. 
D.~Kharzeev is supported by the U.S. Department of Energy, Office of Science, Office of Nuclear Physics, Grants No. DE-FG88ER41450 and DE-SC0012704 and by the U.S. Department of Energy, Office of Science, National Quantum Information Science Research Centers, Co-design Center for Quantum Advantage (C2QA) under Contract No.DE-SC0012704.
The work of C.J.~Naïm and A. Deshpande are supported by the U.S. Department of Energy, DE‐FG02‐05ER41372
The work of Z.~Tu is supported by the U.S. Department of Energy under Award DE-SC0012704 and the BNL Laboratory Directed Research and Development (LDRD) 23-050 project.
\end{acknowledgments}
\clearpage

\bibliography{reference}
\end{document}

\preprint{APS/123-QED}
% \title{Entanglement entropy as a probe of the hadronization process}
\title{Supplemental Material: Entanglement as a probe of hadronization}

%\thanks{Corresppnding author: \email{charlesjoseph.naim@stonybrook.edu}}%
\author{Jaydeep Datta}
\email{jaydeep.datta@stonybrook.edu}
\affiliation{Center for Nuclear Frontiers in Nuclear Science, Department of Physics and Astronomy, Stony Brook University, New York 11794-3800, USA}
\author{Abhay~Deshpande}
\email{abhay.deshpande@stonybrook.edu}
\affiliation{Center for Nuclear Frontiers in Nuclear Science, Department of Physics and Astronomy, Stony Brook University, New York 11794-3800, USA}
\affiliation{Department of Physics, Brookhaven National Laboratory, Upton, New York 11973-5000, USA}
\author{Dmitri E.~Kharzeev}
\email{dmitri.kharzeev@stonybrook.edu}
\affiliation{Center for Nuclear Theory, Department of Physics and Astronomy, Stony Brook University, New York 11794-3800, USA}
\affiliation{Energy and Photon Sciences Directorate, Condensed Matter and Materials Sciences Division, Brookhaven National Laboratory, Upton, New York 11973-5000, USA}
\author{Charles Joseph Naïm}
\email{charlesjoseph.naim@stonybrook.edu}
\affiliation{Center for Nuclear Frontiers in Nuclear Science, Department of Physics and Astronomy, Stony Brook University, New York 11794-3800, USA}
\author{Zhoudunming~Tu}
\email{zhoudunming@bnl.gov}
\affiliation{Department of Physics, Brookhaven National Laboratory, Upton, New York 11973-5000, USA}

\date{\today}

\maketitle

\section{Details on the relationship between  $\langle z \rangle$ and $p_{\perp}^{\text{jet}}$}
\label{sec:z_pt}
\begin{figure}[!h]
    \centering
    \includegraphics[scale=0.5]{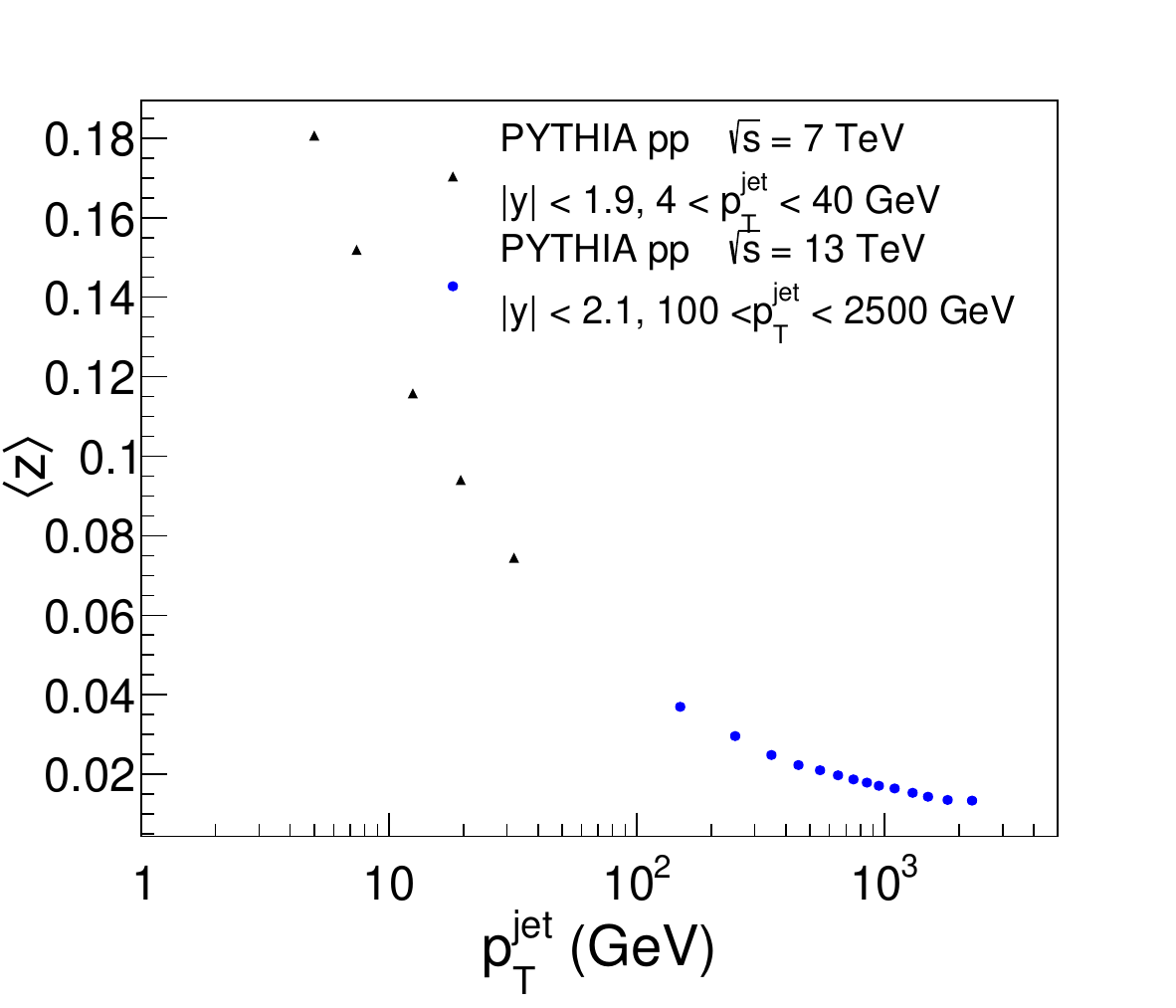}
    \caption{The $\langle z \rangle$ and $p_{\perp}^{\text{jet}}$ correlation at $\sqrt{s} = 7$ (black points) and $\sqrt{s} = 13 \, \text{TeV}$ (blue points) in $pp$ collisions using PYTHIA~8 event generator.}
    \label{fig:5z_vs_pTjet}
\end{figure}

We have re-expressed the ATLAS data at $\sqrt{s} = 7 \, \text{TeV}$ and $\sqrt{s} = 13 \, \text{TeV}$ as a function of $\langle z \rangle$, the most natural variable for comparing the data to our model in Eq.~(1). For each experimental bin in $p_{\perp}^{\text{jet}}$, an average value of $z$ was determined using a PYTHIA 8 simulation for each center-of-the mass energy. This proton-proton simulation includes the hard QCD processes $gg \to q\bar{q}$, $gg \to gg$ and $qg \to qg$, with jets reconstructed using the anti-$k_T$ algorithm with a radius parameter $R = 0.4$. At large $p_{\perp}^{\text{jet}} \gg 1$ GeV, the Compton scattering process $qg \to qg$ becomes dominant. For each experimental $p_{\perp}^{\text{jet}}$ bin, the distribution of the $z$ variable within the jet is determined, and the mean value $\langle z \rangle$ is computed. Figure~\ref{fig:5z_vs_pTjet} illustrates the correlation between $\langle z \rangle$ and $p_{\perp}^{\text{jet}}$ at $\sqrt{s} = 7 \, \text{TeV}$ and $\sqrt{s} = 13 \, \text{TeV}$, within the same kinematic phase space, as compared to ATLAS data (see Tab.~I). 
The ATLAS experiment published the correlation between $\langle z \rangle$ and $p_{\perp}^{\text{jet}}$ at $\sqrt{s} = 7 \, \text{TeV}$ for $ 20 < p_{\perp}^{\text{jet}} < 500$ GeV for $|y|<1.2$ using various physics generators, including PYTHIA \cite{ATLAS:2011myc}. Our calculations show a good agreement with theirs. 
\section{Impact of the $p_{\perp}^{\text{hadrons}}$ threshold on the entropy:}
\label{sec:pT_treshold}
In ATLAS data, the $p_{\perp}^{\text{hadrons}}$ threshold for charged hadrons is 0.3 $\mathrm{GeV}$ at $\sqrt{s} = 7$ TeV and 0.5 $\mathrm{GeV}$ at $\sqrt{s} = 13$ TeV, reducing observed charged hadron multiplicity and affecting entropy. Figure 2 in the main manuscript shows ATLAS data at 7 TeV and a rescaled version using PYTHIA without a $p_{\perp}^{\text{hadrons}}$ cutoff, revealing a 10\% entropy increase, especially at low $p_{\perp}^{\text{jet}}$. At 13 TeV, the cutoff effect diminishes at high $p_{\perp}^{\text{jet}}$. Standard fragmentation functions lack accuracy at low $p_{\perp}^{\text{hadrons}}$, where transverse momentum dependent FF are necessary to capture hadronization in the non-perturbative regime.
\paragraph{Determination of $z_{\text{min}}$:}
To set the lower bound $z_{\text{min}}$ in Eq. (1), we selected $z_{\text{min}}$ such that, for each $p_{\perp}^{\text{jet}}$ bin, the integral of the $z$ distribution equals 50\% for $z > z_{\text{min}}$. This captures the main fragmentation contributions to the charged-hadron multiplicity but also highlights the difficulty of describing soft particles at small $z \lesssim 10^{-2}$ in current fragmentation function extractions.
\section{Determination of the average parton jet fraction}
The fraction of partons initiating jet formation in pp collisions at ATLAS experiments and ep collisions at ePIC experiment was determined using PYTHIA 8 event generator. Only gluons and $u$ and $d$ (anti)quarks were considered. An average estimate of each contribution was obtained by fitting a constant function over the range of jet $p_{\perp}^{\text{jet}}$ considered. These values are summarized in Table \ref{tab:jetfraction}.
\begin{table}[h!]
    \centering
    \begin{tabular}{|c|c|c|}
        \hline
        Experiment &  Gluon jet & Quark jet\\ \hline
        ePIC  & 0.30 & 0.70 \\ \hline
        ATLAS 7 TeV  & 0.89 & 0.11 \\ \hline
        ATLAS 13 TeV & 0.60 & 0.40 \\ \hline
    \end{tabular}
    \caption{\label{tab:jetfraction} Average value of gluon and quark jets fraction calculated using PYTHIA 8 event generator.}
    \label{tab:jetfraction}
\end{table}
\section{Comparison between data and other Fragmentation Functions}
\label{sec:FF_choice}
\begin{figure*}
    \centering
    \includegraphics[scale=0.9]{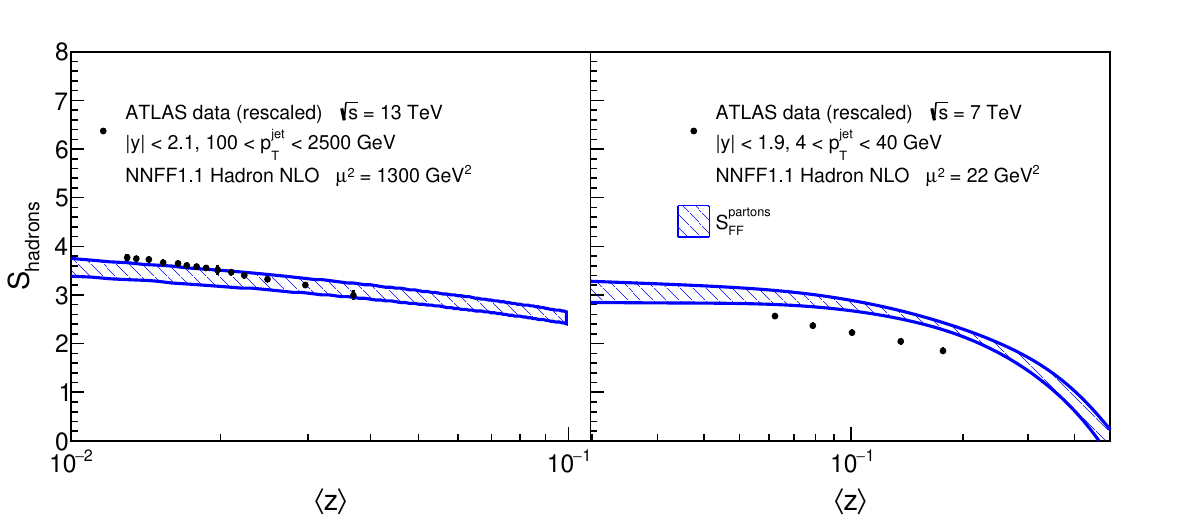}
\caption{The entropy $S_{\text{hadrons}}$ as a function of $\left\langle z \right\rangle$ for $S_{\text{FF}}^{\text{partons}}$ — incorporating gluons, $u$-(anti)quarks, and $d$-(anti)quarks — is shown using NNFF1.1 fragmentation functions at NLO for $\mu^2 = 1300$ GeV$^2$, compared with ATLAS data at $\sqrt{s} = 13$ TeV \cite{ATLAS:2019rqw} (left). Additionally, the results at $\mu^2 = 22$ GeV$^2$ are compared with ATLAS data at $\sqrt{s} = 7$ TeV \cite{ATLAS:2011eid} (right). The uncertainties are calculated at the 1$\sigma$ level. The total entropy $S_{\text{FF}}^{\text{partons}}$ is derived from the sum of the individual entropies of each parton, with each contribution normalized by the average fraction of jets produced by that parton from a PYTHIA simulation.}
\label{fig:NNFF11_ATLAS_data}
\end{figure*}

The ATLAS data at $\sqrt{s} = 7 \, \text{TeV}$ and $\sqrt{s} = 13 \, \text{TeV}$ are also compared to the NNFF1.1 fragmentation function at the NLO in addition to JAM. The results show slightly a different behavior where the predictions from JAM and NNFF1.1 diverge for higher $\langle z \rangle$, as illustrated in Fig.~\ref{fig:NNFF11_ATLAS_data}.

The JAM fragmentation functions \cite{Moffat:2021dji} are derived from a global QCD analysis using a broad dataset, including SIDIS and $e^{+}e^{-}$ annihilation emphasizing uncertainty quantification with extensive replicas. In contrast, NNFF1.1 \cite{Bertone:2018ecm} uses a neural network approach for a flexible, non-parametric FF representation. This methodological difference likely explains observed discrepancies between JAM and NNFF1.1 FFs. NNFF1.1 provides combined $h^+$ and $h^-$ contributions, and we calculated their average values; JAM provides $h^+$ FFs directly, and using charge conjugation, we derived an averaged FF to relate $h^+$ and $h^-$.\\
\section{Energy scale dependence of the Fragmentation Functions}
\label{sec:scale_dependance}
In this subsection, we examine the systematic variation of different hard scale values used by the fragmentation function. With inputs from PYTHIA simulation, the $p_{\perp}^{\text{hadron}}$ is typically from a few $\mathrm{GeV}$ to 10 GeV depending on the jet transverse momentum. To be conservative, we scan the values between 2 to 50 GeV, where our default values are 36 and 4.7 GeV for 13 TeV and 7 TeV data, respectively.

For ATLAS data at $\sqrt{s} = 7 \, \text{TeV}$, JAM shows $\sim$ 5\% variation, NNFF1.1 varies $\sim$ 20\% for gluons and $u$-quarks, and $\sim$ 60\% for $d$-quarks. At $\sqrt{s} = 13 \, \text{TeV}$, entropy changes by 10\%, with a scale at $\approx 5$ GeV. Consequently, despite extensive scale variations, our model remains stable and are within the systematic uncertainty of the FF models.  
\bibliography{reference}